# Switchable anomalous Hall effect by selective mirror symmetry breaking in a kagome magnet GdMn$_6$Ge$_6$

Zicheng Tao,[1, †] Tianye Yu,[2, †] Jianyang Ding,[3, 4 †] Zhicheng Jiang,[4] Zhenhai Yu,[1] Wei Xia,[1,7] Xia Wang,[1, 5] Xuerong Liu,[1,6] Yulin Chen,[7, 8] Dawei Shen,[4,*] Yan Sun,[2,*] and Yanfeng Guo[1, 7, *]

[1]School of Physical Science and Technology, ShanghaiTech University, Shanghai 201210, China

[2]Shenyang National Laboratory for Materials Science, Institute of Metal Research, Chinese Academy of Sciences, Shenyang 110016, China

[3]National Key Laboratory of Materials for Integrated Circuits, Shanghai Institute of Microsystem and Information Technology, Chinese Academy of Sciences, Shanghai 200050, China

[4]National Synchrotron Radiation Laboratory and School of Nuclear Science and Technology, University of Science and Technology of China, Hefei, 230026, China

[5]Analytical Instrumentation Center, School of Physical Science and Technology, ShanghaiTech University, Shanghai 201210, China

[6]Center for Transformative Science, ShanghaiTech University, Shanghai 201210, China

[7]ShanghaiTech Laboratory for Topological Physics, ShanghaiTech University, Shanghai 201210, China

[8]Clarendon Laboratory, Department of Physics, University of Oxford, Oxford OX1 3PU, United Kingdom

†These authors contributed equally to this work:
Zicheng Tao, Tianye Yu, and Jianyang Ding.

*Correspondence:

*dwshen@ustc.edu.cn

*sunyan@imr.ac.cn

*guoyf@shanghaitech.edu.cn

**The crystal symmetry plays a pivotal role in protecting the nontrivial electronic states in a topological phase. Manipulation of the crystal symmetry and hence the nontrivial topological states would serve as a fertile ground to explore exotic topological properties. Combining experimental and theoretical investigations, we demonstrate herein the flexible switch of nontrivial topological states in the single phase of kagome magnet GdMn$_6$Ge$_6$. The intrinsic anomalous Hall effect caused by distinct Berry curvatures along different crystallographic directions is realized through selectively breaking the mirror symmetries in these directions by external magnetic field, which is fully supported by the first-principles calculations. Our results set an explicit example demonstrating the strong correlation between structure symmetry and nontrivial topological states, as well as the switchable topological properties in a single magnetic topological phase.**

**Introduction**

Symmetry and topology represent two basic notations in condensed matter physics for understanding the physical properties of solids. In topological physics, the diverse topological phases can be classified by using topological invariants which are protected by specific crystal symmetries [1-13]. Among various symmetries, the mirror symmetry, one type of spatial symmetry involving reflection across a mirror plane, plays a crucial role in determining the electronic band structure topology and hence the physical properties [14-22]. Topological nodal line semimetals (TNLSMs), which generally serve as precursors for many other topological semimetals or topological insulators (TIs), provide an explicit example demonstrating the importance of mirror symmetry [23-31]. In TNLSMs, the formation of topological nodal line or nodal ring often requires the combination of spatial inversion and time-reversal symmetries. However, the topological nodal line or nodal ring against spin-orbit coupling (SOC) also need the protection from one or more spatial symmetries, especially the mirror or non-symmorphic symmetry. In theory, when a mirror symmetry operation commutes with the Hamiltonian of a topological phase, nodal line can be generated when the SOC is absent. In such mirror symmetry protected TNLSMs, the spin degeneracy can be lifted depending on the strength of SOC and the atomic orbitals that constitute the electronic states. This lift of spin degeneracy then can lead to the evolution of a Dirac nodal line semimetal into various topological phases, such as the Weyl nodal line semimetals (PbTaSe$_2$ [17]), Weyl semimetals (TaAs [14]), or TIs (CaP$_3$ [31]).

The intrinsic anomalous Hall effect (AHE) is suggested to arise from a pseudo-magnetic field caused by the momentum integrated Berry curvature that is initially related to band structure topology and topological invariants. The success of Berry curvature physics therefore provides a general classification of band topology [32-38]. Previous theoretical work indicates that symmetry-protected band degeneracies in the non-relativistic electronic band structure, such as mirror symmetry protected topological nodal line, can produce large intrinsic AHE [39], which provides an opportunity to investigate the correlation between mirror symmetry and nontrivial topological states through measuring the AHE. However, experimental verifications of the contribution of mirror symmetry to the AHE in real materials remain very rare.

The hexagonal $RX_6Y_6$ ($R$-166, $R$ = rare earth, X = Mn, V, and Y = Sn, Ge), which naturally host Dirac cones, flat bands, and van Hove singularities in the electronic band structure, exhibit strong magnetic interactions between 3$d$ Mn and localized 4$f$ $R$ ions and diverse magnetism. They have been conceived as an excellent platform for the realization of quantum anomalous Hall effect (QAHE) and therefore attracted considerable research interest [40-49]. The $R$-166 compounds mostly crystallize into the $P6/mmm$ space group (No. 191) [50-52], which hosts multiple mirror planes, such as $M_x$, $M_y$ and $M_z$, as illustrated in Figs. S1(a)-(c) in the supplementary information (SI). Therefore, they can serve as an excellent platform to study the correlation between the mirror symmetry and the intrinsic AHE.

In this work, we have elaborated the principle of mirror symmetry breaking induced AHE by using the kagome magnet $GdMn_6Ge_6$ as an explicit example. Combining the angle-resolved photoemission spectroscopy (APRES), magnetotransport measurements and first-principles calculations, we have unveiled very large intrinsic anomalous Hall conductivities (AHCs) of 1263 $\Omega^{-1}$ cm$^{-1}$, 1163 $\Omega^{-1}$ cm$^{-1}$ and 374 $\Omega^{-1}$ cm$^{-1}$ for the magnetic field $H//x$, $H//y$ and $H//z$, respectively. The significant AHE is tightly related to the tunable Berry curvature by breaking the mirror symmetries along different crystallographic directions by the external magnetic field.

The details for crystal growth, crystal structure characterizations, calculated magnetic structures and the corresponding band structures, magnetotransport, ARPES measurements, and first-principles calculations of the $GdMn_6Ge_6$ crystals are presented in the SI which includes Figs. S1-S11, Table S1 and Refs. [53-64].

**Results and discussions**

The principle of mirror symmetry breaking induced AHE with varying the direction of external magnetic field is schematically illustrated in Fig. 1. When considering the SOC, the topological nodal ring protected by the mirror symmetry perpendicular to the magnetic field direction can be maintained, while other nodal rings will be destroyed. For example, when the magnetic field is along the *z* axis, only the red nodal ring protected by $M_z$ can be maintained, which does not contribute to the AHE. While the blue and green nodal lines protected by $M_x$ and $M_y$ are destroyed, thus contributing to the AHE. When the magnetic field deviates from all the *x*, *y* and *z* axis, all three nodal rings are destroyed and can contribute to the AHE. Consequently, when magnetic field along the *z* axis, the magnitude of AHC at the energy position of the nodal ring is smaller than that for the magnetic field deviating from the *x*, *y* and *z* axis.

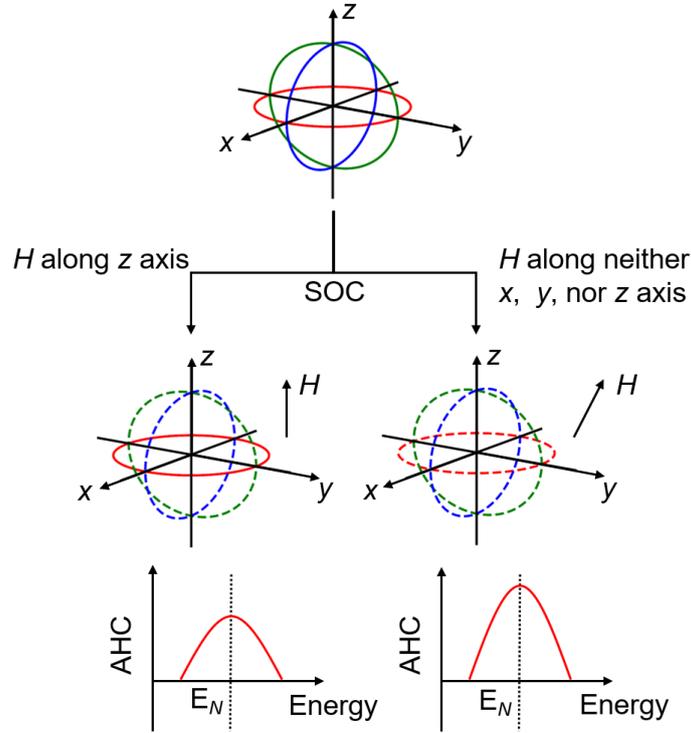

**Fig. 1.** Schematic illustration of mirror symmetry breaking induced AHE with different magnetic field directions. The mirror symmetry protected nodal rings of bands with the same spin are shown by solid green, red and blue lines, which are protected by the mirror planes $M_x$, $M_y$ and $M_z$ (corresponding to the $k_x = 0$, $k_y = 0$ and $k_z = 0$ planes), respectively.

$GdMn_6Ge_6$ exhibits a long-range ferromagnetic order with the Curie temperature

even reaching ~ 490 K [60]. In order to investigate the electronic band structure properties of GdMn$_6$Ge$_6$ in its magnetic ground state, we investigate its electronic structure by using ARPES and first-principles calculations. The three-dimensional (3D) Brillouin zone (BZ) with high-symmetry points and projected BZ along $k_z$-axis are presented in Fig. 2(a). Through the periodicity of the $k_z$ photoemission intensity map near $E_F$ (Fig. S11(a) of SI), the inner potential is determined as 19 eV, with photon energies of 95 and 109 eV corresponding to Γ and A points, respectively. Figs. 2(b)-(c) show the Fermi surface of the ΓKM and AHL planes, respectively, and the evolution of the constant energy contours with the binding energy ($E_B$). The Fermi surface exhibits a hexagonal petal-like morphology, similar to other *R*-166 compounds [33,44-47,65-68]. Large enclosed triangular pockets, labeled as *α,* are observed near the K and H points, while an internal pocket denoted as *β* is distinctly present only at the H point. Unlike $R$Mn$_6$Sn$_6$ ($R$ = Sc, Y and lanthanides), the α pocket in GdMn$_6$Ge$_6$ does not gradually shrink into a dot with the increasing of $E_B$, thus failing to form a Dirac cone near $E_F$. Figs. 2(e)-(f) depict the dispersion along Γ-K-M-Γ and A-H-L-A directions ($k$ space path illustrated in Fig. 2(d)), together with the appended calculated bulk bands (the orange solid lines) of the ground state. The detailed calculation process is discussed in the SI. In comparison with the ARPES results, the $E_F$ in the DFT calculations need to be shifted upward of 0.064 eV and renormalized with a factor of 1.2 to achieve a better match to the overall photoemission intensity plot. Importantly, we have identified the presence of a saddle point (SP) at around 30 meV below $E_F$, as a typical character of kagome band structure. The SP with high density of states lies in close proximity to $E_F$ and is present at both M and L points, similar as that in YMn$_6$Sn$_6$, indicating the negligible $k_z$ dispersion. Additionally, the evolution of dispersion near M point (parallel to the Γ-M direction: $k_y$=-0.3-0.3Å) also confirm the existence of the SP (Fig. S11(d)). The stacking plots mentioned above clearly reveal that the electronic band structure can be partitioned by multiple mirror planes, indicating the necessary role of mirror symmetry.

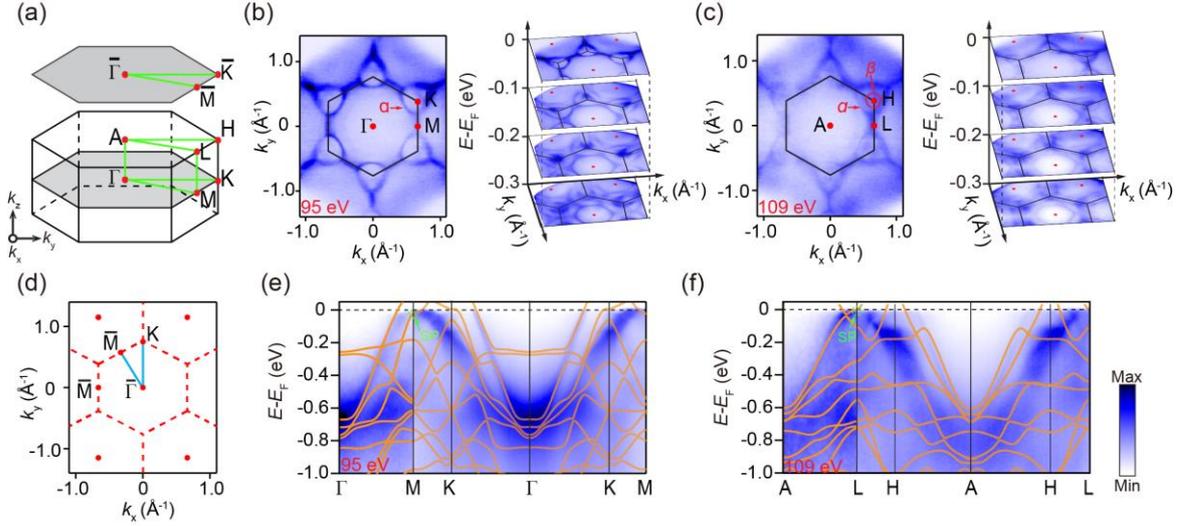

**Fig. 2.** (a) The 3D Brillouin zone of hexagonal GdMn$_6$Ge$_6$ and its projection on the $k_x$-$k_y$ plane. (b)-(c) Left: Fermi surface mapping. Right: stacking plots of constant energy contours from $E_F$ to $E_F$ - 0.3 eV at 95 eV (Γ point) and 109 eV (A point), respectively. (d) Schematic diagram of dispersion along high symmetry point. (e)-(f) Photoelectron intensity plots of the band structure taken along Γ-K-M-Γ and A-H-L-A directions, respectively. The calculated bulk bands are appended by the orange lines.

To yield more insights into the electronic band structure of GdMn$_6$Ge$_6$, the magnetizations and magnetotransport properties were also investigated. The isothermal magnetization $M(H)$ curves and the corresponding longitudinal and transverse resistivity ($\rho_{MR}$ and $\rho_H$) under magnetic field applied along the $x, y$ and $z$ directions are shown in Fig. S5 of SI. The measurement configurations for $\rho_H$ are illustrated in the insets of Figs. 3(d)-3(f). The anomalous Hall resistivity $\rho_{AH}$ for μ$H_0$ along the $x, y$ and $z$ directions are depicted in Figs. 3(a)-3(c), which are extracted from $\rho_H$, as discussed in the SI. The fits to the log($\rho_{AH}$) against the log($\rho_{MR}$) curves above 100 K by using the formula $\rho_{AH} \propto \rho_{MR}^\alpha$ yield the scaling exponents α of 1.9, 1.8 and 1.8 for $H//x$, $H//y$ and $H//z$, respectively, thus excluding the possible source of extrinsic skew-scattering mechanism in which $\rho_{AH} \propto \rho_{MR}$ [69], as shown in Figs. 3(d)-3(f). Alternatively, the intrinsic Berry curvature or by the extrinsic side-jump effect both can account for the AHE in GdMn$_6$Ge$_6$ [70-72]. To trace the true origin, the Tian-Ye-Jin (TYJ) scaling model is used to extract the intrinsic AHC $\sigma_{AH}$ ($\sigma_{AH} = -\rho_{AH}/[(\rho_{AH})^2 + \rho_{MR}^2]$), which is described as

$$\sigma_{AH} = -(\alpha\sigma_{MR0}^{-1} + \beta\sigma_{MR0}^{-2})\sigma_{MR}^2 - b \quad (1),$$

where $\sigma_{MR0} = 1/\rho_{MR0}$ is the residual conductivity. The intrinsic $\sigma_{AH}$ is given by the intercept $b$ [61,73], as illustrated in Fig. 3(g). The linear relation between $\sigma_{AH}$ and $\sigma_{MR}^2$ indicates the intrinsic AHE caused by Berry curvature rather than side-jump effect. It is worth noting that the $\sigma_{AH}$ for $H//x$ and $H//y$ are much larger than that for $H//z$. Besides, the obtained $\sigma_{xy}^A$ are 1263 $\Omega^{-1}$ cm$^{-1}$, 1186 $\Omega^{-1}$ cm$^{-1}$ and 374 $\Omega^{-1}$ cm$^{-1}$ for $H//x$, $H//y$ and $H//z$, respectively, which are much larger than those of other $R$-166 compounds [42,63,74].

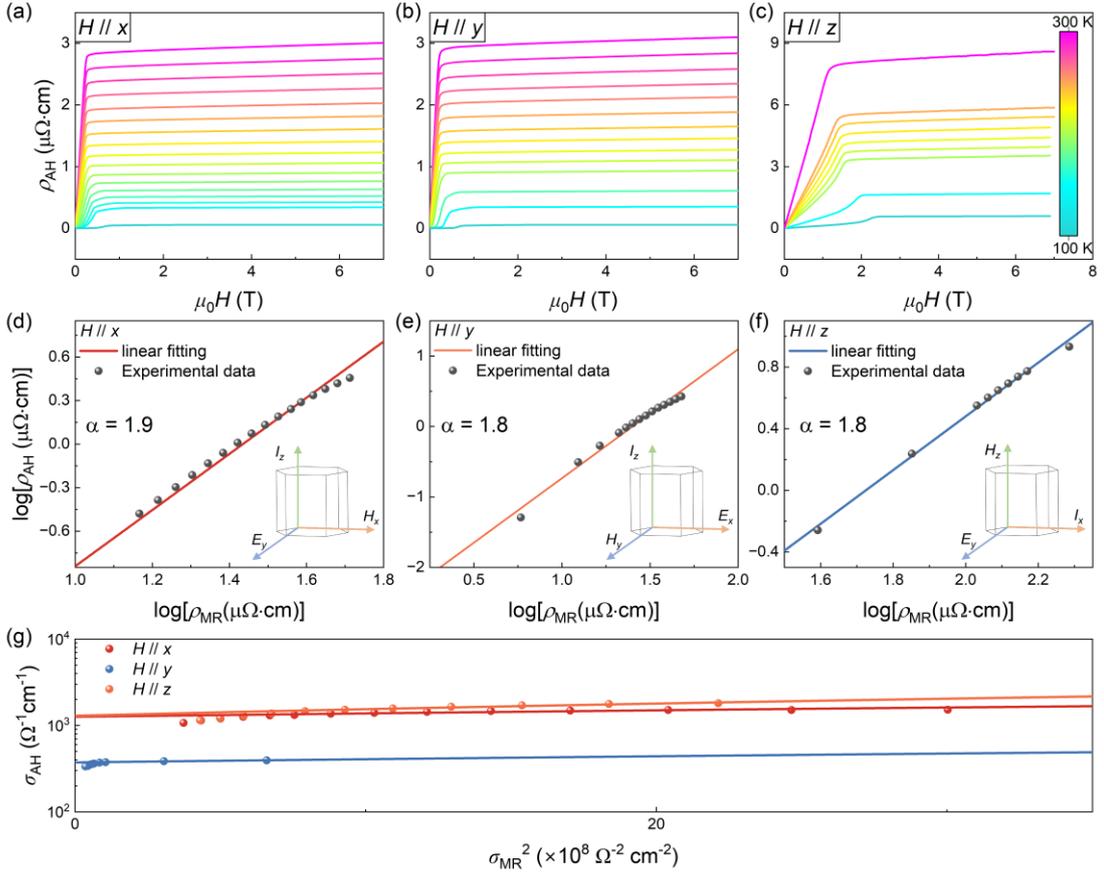

Fig. 3. (a)-(c) The magnetic field dependence of anomalous Hall resistivity $\rho_{AH}$ extracted by using $\rho_H = \rho_{NH} + \rho_{AH} = R_0B + R_S4\pi M$ at various temperatures for $\mu H_0$ along the $x$, $y$ and $z$ directions. (d)-(f) Plot of log($\rho_{AH}$) against log($\rho_{MR}$) for $H//x$, $H//y$ and $H//z$. The lines represent the fitting curve by a linear function. (g) The anomalous Hall conductivity $\sigma_{AH}$ versus $\sigma_{MR}^2$ with fitting lines for $H//x$, $H//y$ and $H//z$, respectively. The insets of (d)-(f) show the measurement configurations for $\rho_{AH}$ of GdMn$_6$Ge$_6$ when $H//x$, $H//y$ and $H//z$, respectively.

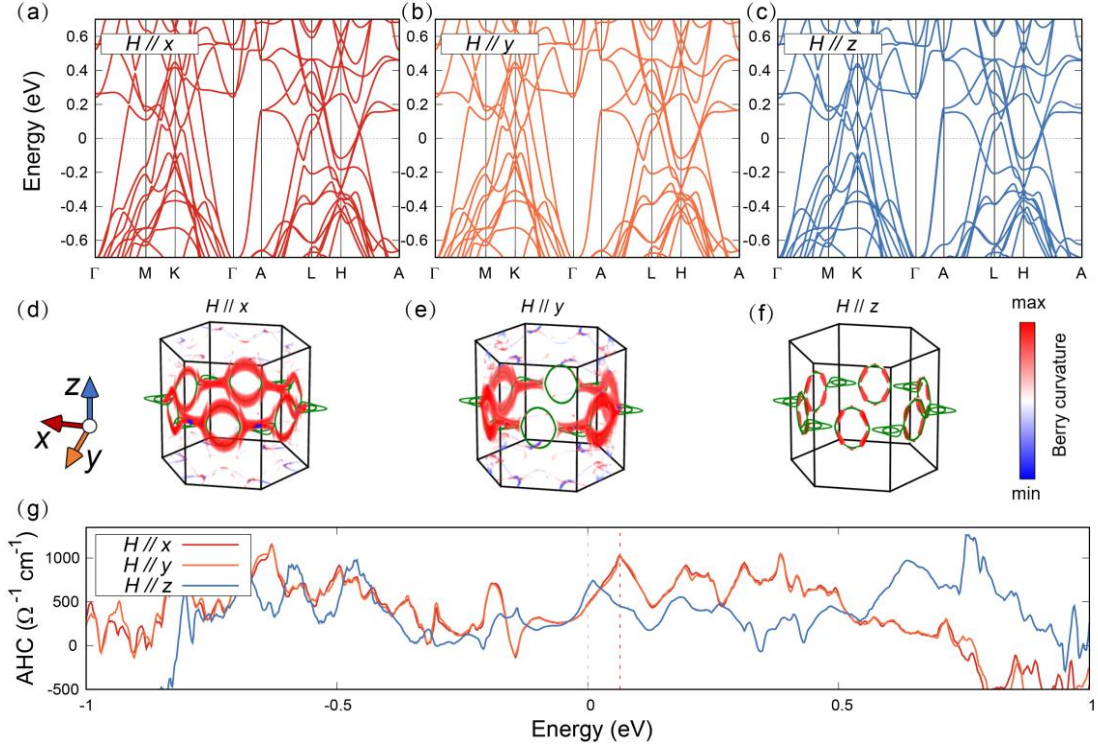

**Fig. 4.** (a)-(c) The band structures of GdMn$_6$Ge$_6$ for $H//x$, $H//y$ and $H//z$, respectively. The corresponding magnetic configurations are shown in Fig. S9 of SI. The Berry curvature of GdMn$_6$Ge$_6$ for (d) $H//x$, (e) $H//y$ and (f) $H//z$, respectively. The blue circles denote the nodal rings near $E_F$, which are corresponding to the gap opening induced Berry curvature. (g) AHC of GdMn$_6$Ge$_6$ for $H//x$, $H//y$ and $H//z$, respectively. The dashed red vertical line denotes $E_F$+0.064 eV.

In order to understand the large intrinsic AHC in GdMn$_6$Ge$_6$, the band structures of GdMn$_6$Ge$_6$ with considering the SOC for $H//x$, $H//y$ and $H//z$ are presented in Figs. 4(a)-4(c). It is apparent that the change of magnetic field direction dos not substantially affect the electronic band structure. The comparison of calculated band structures with the ARPES measurement results shows that the $E_F$ should be lifted about 0.064 eV. For the AHC calculation, the $E_F$ is therefore also lifted in an energy with the same magnitude. Without taking the SOC into account, the calculations indicate that there are nodal rings near $E_F$, as shown in Figs. S10(a)-S10(c) of the SI. As shown in Fig. 4(g), once the SOC is considered, the nodal rings are disrupted with different extents under varied applied magnetic field directions, thus resulting in AHC with different magnitudes. For $H//x$, the nodal rings on both $k_z = 0$ and MKL planes are destroyed, yielding the largest AHC of 1023 $\Omega^{-1}$ cm$^{-1}$. For $H//y$, the nodal rings on the $k_z = 0$ plane are destroyed but two nodal rings on the two MKL planes vertical to $y$ direction are maintained, which cannot contribute to the AHC in such case, as shown in Fig. 4(e). The reduction in AHC is not significant owing to the small amount of residual nodal rings. As a result, the AHC for $H//y$ is 1011 $\Omega^{-1}$ cm$^{-1}$. For $H//z$, the nodal rings on the MKL planes are all destroyed but the nodal rings on the $k_z = 0$ planes are maintained, without contributions to Berry curvature, resulting in a smallest AHC of 454 $\Omega^{-1}$ cm$^{-1}$. The evolution of AHC magnitude when the direction of applied field changes from $x$ to $y$ and then to $z$ shows good agreement with the experimental values of 1263 $\Omega^{-1}$ cm$^{-1}$, 1186 $\Omega^{-1}$ cm$^{-1}$ and 374 $\Omega^{-1}$ cm$^{-1}$ for $H//x$, $H//y$ and $H//z$, respectively, thus exposing the switchable intrinsic AHE in GdMn$_6$Ge$_6$ by external magnetic field.

**Conclusion**

In summary, GeMn$_6$Sn$_6$ hosts multiple distinct mirror symmetries in the electronic band structure along different crystallographic directions, which protect topological nodal rings near the Fermi level, as confirmed by our APRES measurements and first-principles calculations. Through selectively breaking the mirror symmetries along the different crystallographic directions by external magnetic field, the topological nodal rings are disrupted with different extents, thus resulting in different magnitudes of Berry curvature and hence the intrinsic anomalous Hall conductivity. Our work sets an excellent paradigm for the investigation of the correlation between mirror symmetry and anomalous Hall effect, which would be valuable to brush up our knowledge on topological physics. Besides, the switchable anomalous Hall effect in a single topological phase also offers opportunity for practical use in topological devices.

**Acknowledgements**

The authors acknowledge the National Key R&D Program of China (Grants No. 2023YFA1406100, 2021YFB3501503, 2023YFA1406304), the National Nature Science Foundation of China (Grants No. 920651, 52271016, 52188101, 11934017, U2032208), and the Shanghai Science and Technology Innovation Action Plan (Grant No. 21JC1402000). Y.F.G. acknowledges the open research fund of Beijing National Laboratory for Condensed Matter Physics (2023BNLCMPKF002). W.X. thanks the support by the open project from State Key Laboratory of Surface Physics and Department of Physics, Fudan University (Grant No. KF2022-13) and the Shanghai Sailing Program (23YF1426900). Part of this research used Beamline 03U of the Shanghai Synchrotron Radiation Facility, which is supported by $ME^2$ project under contract No. 11227902 from National Natural Science Foundation of China. The authors also thank the support from Analytical Instrumentation Center (#SPST-AIC10112914) and the Double First-Class Initiative Fund of ShanghaiTech University. Part of the numerical calculations in this study were carried out on the ORISE Supercomputer.

# SI

## A. Crystal growth and magnetotransport measurements

Single crystals of GdMn$_6$Ge$_6$ were grown by using the flux method. Starting materials of Gd powders (99.9%, Macklin), Mn pieces (99.9%, Macklin), Ge ingots (99.9%, Macklin) and Sn grains (99.99%, Macklin) were mixed in a molar ratio of 1:6:10:20 and placed into an alumina crucible, which were then sealed into a quartz tube in vacuum. The assembly was heated in a furnace up to 1050 °C within 10 h, kept at that temperature for 10 h, and then slowly cooled down to 700 °C at a temperature decreasing rate of 2 °C/h. The excess flux was removed at this temperature by quickly placing the assembly into a high-speed centrifuge. Hexagonal crystals with the size of 4 mm were left in the alumina crucible. The composition analysis was performed on a Phenom Pro scanning electron microscope (SEM) with the energy dispersive X-ray spectroscopy (EDS). The measured Gd: Mn: Ge is ~ 0.96: 6: 6.66 that is close to 1:6:6, demonstrating the stichometry of the crystals. The single-crystal X-ray diffraction data were collected on a Bruker D8 single-crystal X-ray diffractometer (SXRD) with Mo K$_{\alpha 1}$ ($\lambda$ = 0.71073 Å) at 278 K. The powder XRD (PXRD) was carried on a Bruker D8 Advance powder X-ray diffractometer (PXRD) with Cu K$_{\alpha 1}$ ($\lambda$ = 1.54184 Å) at 298 K. The collected SXRD and PXRD data was analyzed by using Olex2 and GSAS-II softwares, respectively [53,54]. The electrical- and magneto-transport measurements, including the resistivity and Hall effect measurements, were carried out by using a standard Hall bar geometry in a commercial DynaCool Physical Properties Measurement System from Quantum Design. The magnetic susceptibility measurements were measured on a Quantum Design Magnetic Properties Measurement System.

## B. Angle-resolved photoemission spectroscopy (ARPES) measurements

The high-resolution ARPES measurements were performed at the BL03U beamline of the Shanghai Synchrotron Radiation Facility (SSRF) [55,56]. The energy and angular resolutions were set to better than 20 meV and 0.02 Å$^{-1}$, respectively. The light spot size is smaller than 20 μm. Samples were cleaved *in situ*, exposing flat mirror-like (001) surfaces. During measurements, the temperature was kept at $T \sim 15$ K and the pressure was maintained at less than $8 \times 10^{-11}$ Torr. The $k_z$ photoemission mapping used photon energies from 68 to 140 eV.

## C. First-principles calculations

We performed first-principle calculations calculations to study GdMn$_6$Ge$_6$ by using the projector augmented wave method as implemented in Vienna *ab initio* Simulation Package (VASP) [57]. We used the Perdew Burke-Ernzerhof (PBE) exchange correlation functional [58]. The cutoff energy for the plane-wave basis was set to be 450 eV. In order to obtain a convergent result when the applied magnetic field is used, the orientations of magnetic moments of atoms Gd and Mn were fixed in our calculations. We used WANNIER90 to obtain the tight binding model Hamiltonian and then calculated the intrinsic anomalous Hall conductivity (AHC) [59].

## D. Crystal structure and X-ray diffraction data

GdMn$_6$Ge$_6$ has the same crystal structure with other hexagonal $R$X$_6$Y$_6$ ($R$-166, $R$ = rare earth, X = Mn, V, and Y = Sn, Ge) compounds, with the lattice parameters $a = b$ = 5.246 Å, $c$ = 8.1968 Å, and $α = β = 90°$, $γ = 120°$, as illustrated in Figs. S1(a)-S1(b).

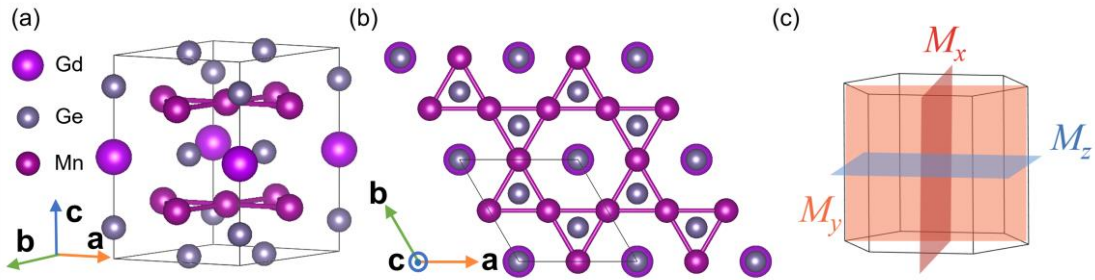

**Fig. S1.** (a)-(b) Schematic views of the GdMn$_6$Ge$_6$ crystal structure from different perspectives. (c) Schematic diagram of multiple mirror planes in the crystal structure of the R166 family.

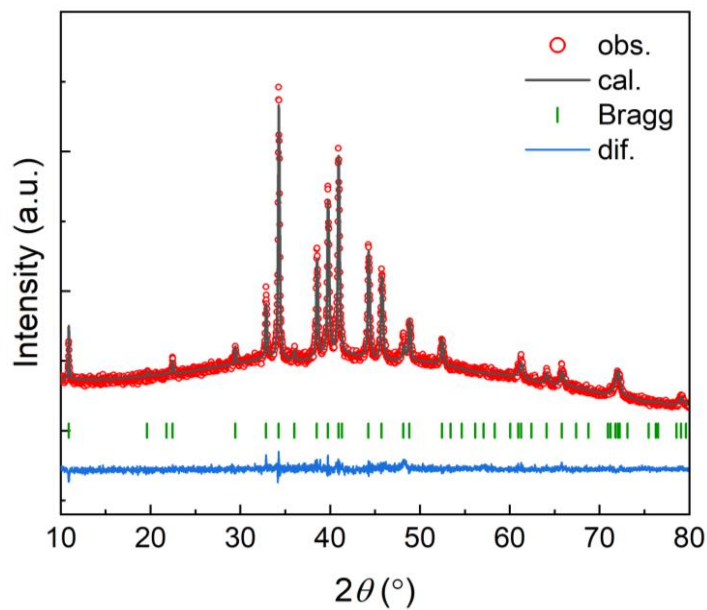

**Fig. S2.** Powder XRD diffraction patterns of GdMn$_6$Ge$_6$ measured at 298 K.

**Table. S1.** The refined parameters from PXRD measurements.

| Site | Wyckoff | Symmetry | x | y | z | Occupation | Uiso |
|---|---|---|---|---|---|---|---|
| Gd1 | 1a | 6/mmm | 1 | 1 | 1/2 | 1 | 0.006 |
| Ge1 | 2d | -6m2 | 2/3 | 1/3 | 1/2 | 1 | 0.004 |
| Ge2 | 2e | 6mm | 1 | 1 | 0.14859 | 1 | 0.006 |
| Ge3 | 2c | -6m2 | 2/3 | 1/3 | 0 | 1 | 0.007 |
| Mn1 | 6i | mm2 | 1/2 | 1/2 | 0.2485 | 1 | 0.006 |

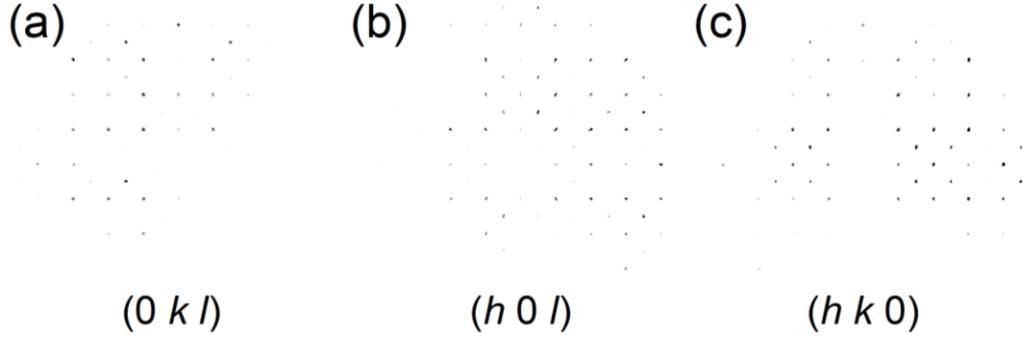

**Fig. S3.** (a)-(c) The diffraction patterns in the reciprocal space along ($0kl$), ($h0l$), and ($hk0$) directions.

This structure processes three primary mirror symmetries: $M_x$ (passes through the principal axis and is perpendicular to the edges), $M_y$ (contains the principal axis but is perpendicular to the side faces), and $M_z$ (parallel to the top and bottom surfaces and is located at half the height of the structure), as shown in Fig. S1(c). PXRD refinement results confirm that GdMn$_6$Ge$_6$ single crystal crystallizes into the hexagonal HfFe$_6$Ge$_6$-type structure with space group $P6/mmm$ (No. 191), as depicted in Fig. S2. The refined atomic parameters are summarized in Table. S1.

The perfect diffraction patterns in the reciprocal space without any other miscellaneous points, seen in Figs. S3(a)-3(c) demonstrate the pure phase and high quality of the crystal used in this study.

### E. Magnetotransport data and analysis

Fig. S4(a) shows the temperature dependent longitudinal resistivity $\rho(T)$ of GdMn$_6$Ge$_6$ with the electrical current along the crystallographic $a$-axis and $c$-axis, which unveils the metallic nature of GdMn$_6$Ge$_6$ within 2-300 K. The ratio of $\rho(T)$ for $I//c$ and $I//a$ is close to 3, signifying the clear anisotropy and indicating the system as a 3D system. The temperature of magnetization $M(T)$ curves in Fig. S4(b) for GdMn$_6$Ge$_6$ under an external magnetic field $\mu_0 H$ of 0.1 T along the $x$, $y$ and $z$ directions show the enhancement of M above 220 K, which is consistent with previous study [60]. Previous work suggested that GdMn$_6$Ge$_6$ manifests long-range ferromagnetic order with the Curie temperature reaching 490 K and has an easy plane parallel to the Mn-based kagome lattice (the *ab* plane).

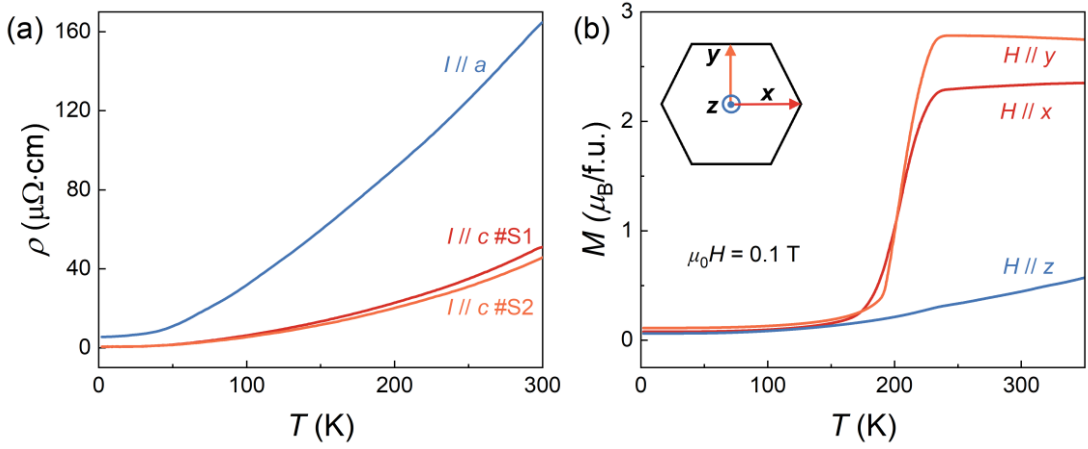

**Fig. S4.** (a) Temperature dependent longitudinal resistivity $\rho$ of GdMn$_6$Ge$_6$ with the current along the *a*-axis and *c*-axis, respectively. (b) Temperature dependence of magnetization along the *x, y* and *z* directions at $\mu_0H = 0.1$ T. Inset: crystallographic orientations.

Within the measurement temperature range, the magnetic field dependent longitudinal resistivity $\rho_{MR}$ for $\mu_0H$ along the *x, y* and *z* directions does not exhibit significant large magnetoresistance, which could be attributed to the magnetic scattering within the material, as shown in Figs. S5(a)-S5(c). Figs. S5(d)-S5(f) show that the *M(H)* curves feature a saturation plateau at relatively large magnetic field with *H//z* while the saturation plateau occurs at much lower field for *H//x* and *H//y*, indicating again that the *ab* plane is the easy plane. In this system, magnetism is primarily attributed to the strong exchange among Mn spins, while the spins of the rare earth ions align antiparallel to those of Mn. However, at high temperatures, the magnetic moment of rare earth is relatively small due to spin reorientation, which may explain the increase of magnetization with raising the temperature. The behavior described above highly resembles that of TbMn$_6$Ge$_6$ with the flat spiral magnetic structure. It is thus speculated that GdMn$_6$Ge$_6$ also feature a complex spiral magnetic structure. Unfortunately, due to the huge absorption cross section of the Gd nucleus, a neutron diffraction study of GdMn$_6$Ge$_6$ is very difficult and the magnetic structure therefore remains unknown.

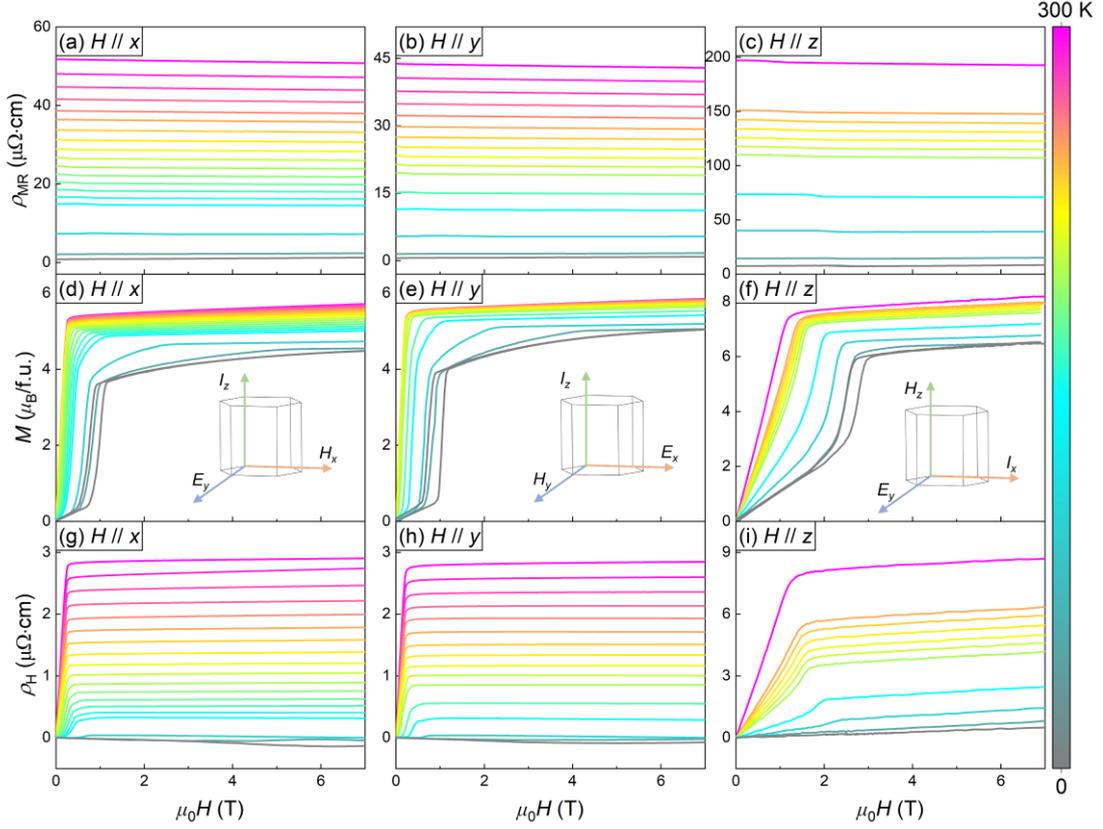

**Fig. S5.** (a)-(c) The magnetic field dependence of (a)-(c) the longitudinal resistivity $\rho_{MR}$, (d)-(f) magnetizations $M$, and (g)-(i) the transverse resistivity $\rho_H$ at various temperatures for $\mu_0 H$ along the $x$, $y$ and $z$ directions. The insets of (d), (e) and (f) show the measurement configurations for $\rho_H$ of GdMn$_6$Ge$_6$ when $H//x$, $H//y$ and $H//z$, respectively.

The $\rho_H$ of GdMn$_6$Ge$_6$ under magnetic field applied along the $x$, $y$ and $z$ directions follow the same trend of the $M(H)$ in the whole temperature range, showing a large anomalous Hall effect (AHE) above 100 K, as displayed in Figs. S5(g)-S5(i). In general, $\rho_H$ can be expressed as

$$\rho_H = \rho_{NH} + \rho_{AH} = R_0 B + R_S 4\pi M \quad (S1),$$

where $\rho_{NH}$ and $\rho_{AH}$ represent ordinary and anomalous Hall resistivity with $R_0$ and $R_S$ denoting ordinary and anomalous Hall coefficient, respectively [61,62]. The anomalous Hall resistivity can be extracted by Eq. S1, which is known to be originated from either intrinsic or extrinsic contributions, respectively. It is obvious that $R_S$ is not zero in GdMn$_6$Ge$_6$, as shown in Figs. S5(g)-S5(i). The anomalous Hall resistivity $\rho_{AH}$, which obviously dominates the total Hall resistivity, is derived by subtracting $\rho_{NH}$ from $\rho_H$. It is noteworthy that the $R_0$ of GdMn$_6$Ge$_6$ is positive for $H//z$, while is negative for both $H//x$ and $H//y$, as shown in Fig. S6, indicating that there is an anisotropic carrier type with different magnetic field directions in GdMn$_6$Ge$_6$, which is similar to that in TbMn$_6$Ge$_6$ and YMn$_6$Sn$_6$ [63,64].

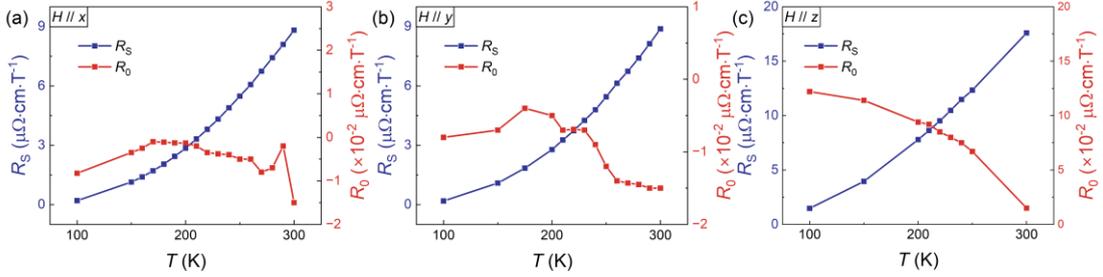

**Fig. S6.** (a)-(c) Ordinary and anomalous Hall coefficients $R_0$ (right axis) and $R_S$ (left axis), as a function of $T$, under different field directions for GdMn$_6$Ge$_6$, respectively.

The ordinary and anomalous Hall coefficients $R_0$ and $R_S$ as a function of $T$ for $H//x$, $H//y$ and $H//z$ are plotted in Fig. S6(a)-S6(c), respectively. $R_S$ is hundreds of times larger than $R_0$ in the measured temperature range. The obtained $R_S$ from the fitting of Eq. S1 in the main text is reliable while $R_0$ has a large error bar, so we can only ascertain the sign of $R_0$ while cannot determine its exact value.

## F. The calculated band structures

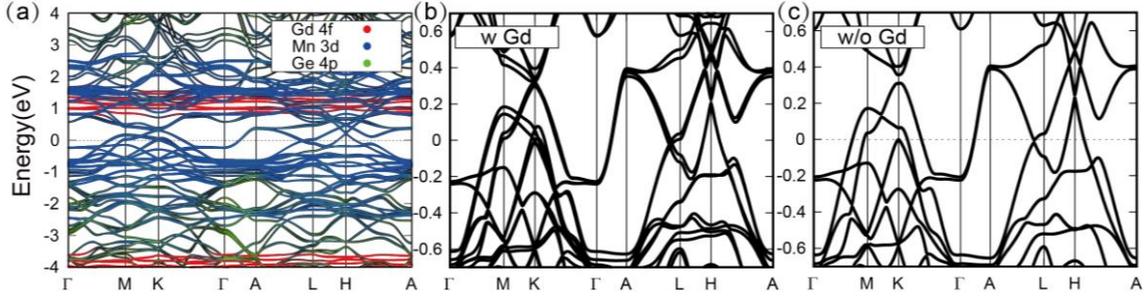

**Fig. S7.** (a) The fat bands of GdMn$_6$Ge$_6$ (the corresponding magnetic configuration shown in Fig. S8(i)). The band structures of GdMn$_6$Ge$_6$ obtained (b) with and (c) without considering the magnetic moment of atom Gd.

In Fig. S7(a), we present the fat bands of GdMn$_6$Ge$_6$ through DFT calculations, with the corresponding magnetic configuration shown in Fig. S8(i). We found that the bands near $E_F$ are predominantly contributed by Mn $3d$ orbitals. We zoom in the energy scale to -0.7 eV to 0.7 eV in Fig. S7(b). In order to illustrate the influence of Gd magnetic moment on the band structure, we plot the band structure without considering the magnetic moment of Gd in Fig. S7(c). Comparing Fig. S7(b) and Fig. S7(c), we find that the magnetic moment of Gd equivalently provide a Zeeman field which split the bands slightly. As a result, we use the bands in Fig. S7(c) to match the ARPES measurements.

## G. The magnetic configurations for *H*//*x*, *H*//*y* and *H*//*z*

To determine the magnetic configuration of GdMn$_6$Ge$_6$ for *H*//*x*, we calculated the band structures of several possible magnetic configurations. The magnetic configuration shown in Fig. S8(a) was found to be in good agreement with the experimental results for *H*//*x*, which is calculated specifically for this study. Figures. S9(a)-S9(f) depict the magnetic configurations of GdMn$_6$Ge$_6$ for *H*//*x*, *H*//*y* and *H*//*z* in this work.

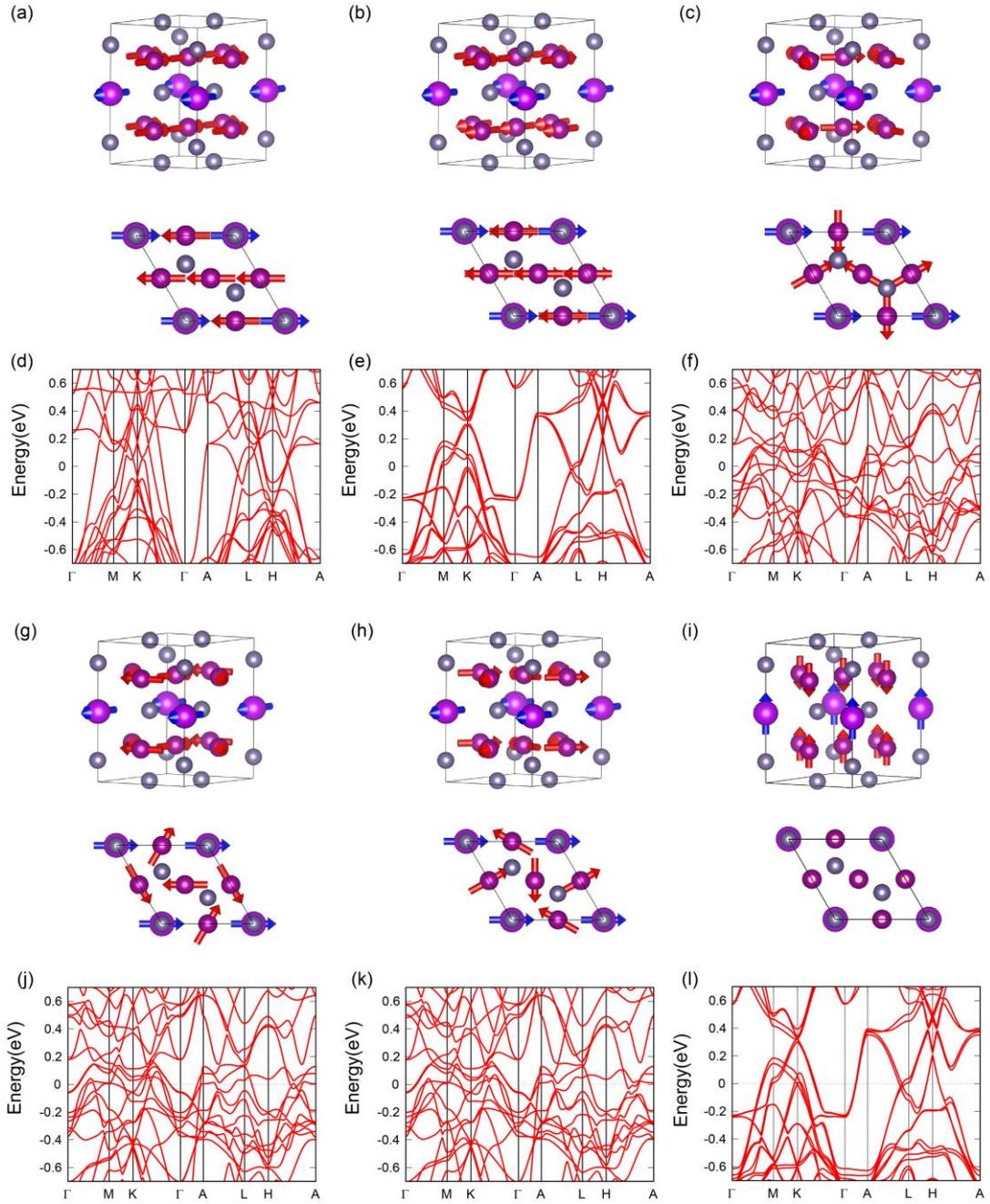

**Fig. S8.** (a)-(c) and (g)-(h) The possible magnetic configurations of GdMn$_6$Ge$_6$ for $H//x$. (d)-(f) and (j)-(k) The corresponding band structures. (i) The magnetic configurations of GdMn$_6$Ge$_6$ for matching the ARPES measurements. (l) The corresponding band structures.

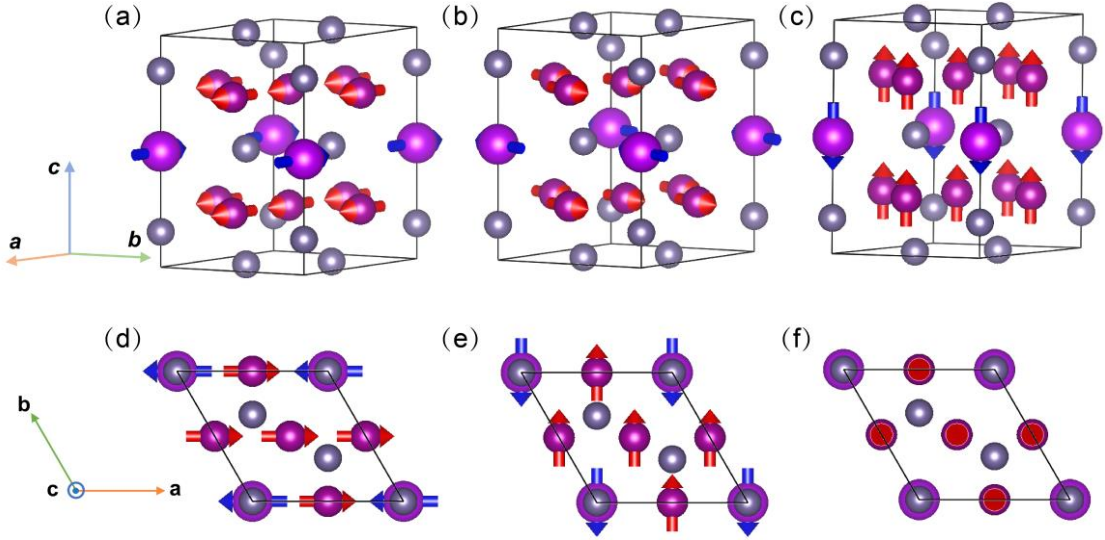

**Fig. S9.** (a)-(c) The magnetic configurations of GdMn$_6$Ge$_6$ for $H//x$, $H//y$ and $H//z$, respectively. (d)-(f)The corresponding top views for $H//x$, $H//y$ and $H//z$.

## H. The topological nodal ring of GdMn$_6$Ge$_6$

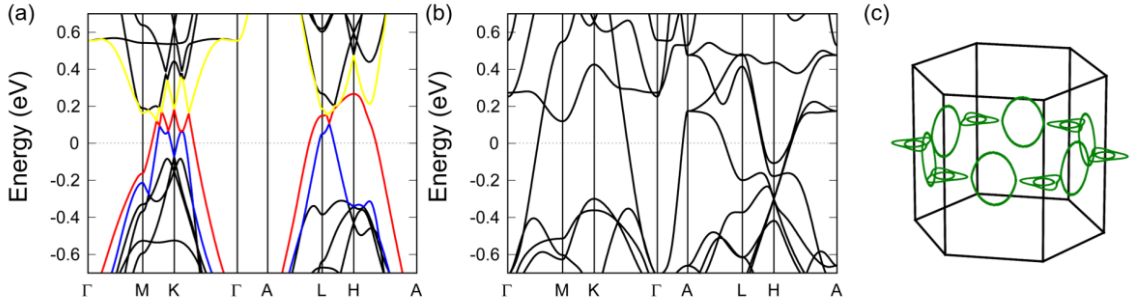

**Fig. S10.** (a)-(b) The position of nodal ring in the electronic band structure. Left: spin-up band structure. Right: spin-down band structure.

The position of nodal ring in the energy band is depicted in Figs. S10(a)-S10(b), which show the nodal rings on the $k_z = 0$ plane, formed by the blue and red bands, and the nodal rings on the MKL planes, formed by red and yellow bands, respectively. When spin-orbit coupling (SOC) is neglected, the nodal rings are present on the $k_z = 0$ and MKL planes, as shown in Fig. S10(c).

# I. ARPES data

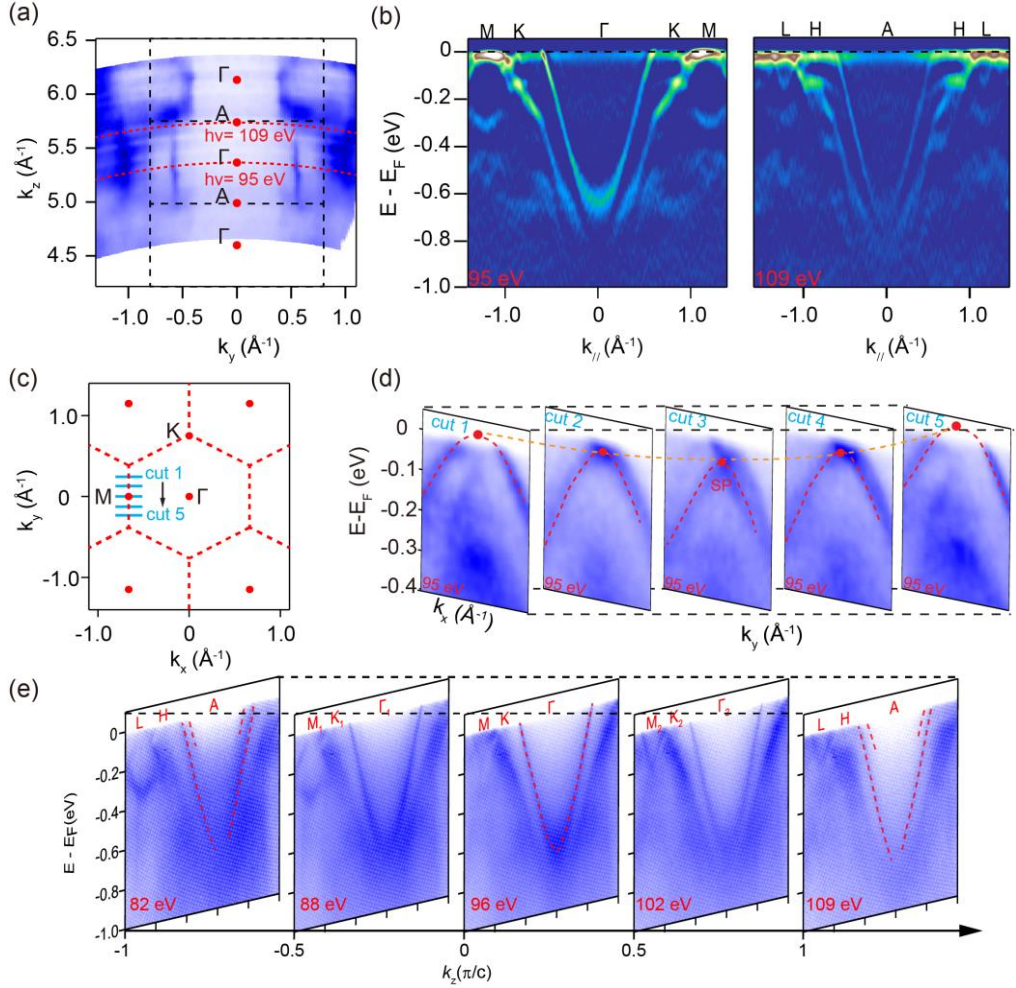

**Fig. S11.** (a) $k_y$-$k_z$ photoemission intensity map at $E_F$ taken with the photon energies ranging from 68 to 140 eV and within an energy window of ± 10 meV. (b) The second-derivative plots of the dispersion along Γ-K-M and A-H-L direction. (c) Schematic diagram of the path of dispersion in the $k$-space near M point. (d) Photoemission intensity plot along cut1-cut5 corresponding to (c). (e) ARPES spectra along A-H-L taken with different photon energies.

The $k_z$ dispersion periodicity, as illustrated in Fig. S11(a), allows for the determination of the photon energies corresponding to the high symmetry points at 95 and 109 eV, respectively. The second-derivative plots displayed in Fig. S11(b) show the band details more clearly. The dispersion along the cut1 to cut5 paths in Fig. S11(c) is presented in Fig. S11(d). The evolution of the top of the hole band deviating from the M point confirms the existence of the SP at the M point. In addition, we mapped out the ARPES spectra along A-H-L with photon energy from 82 eV to 109 eV, shown in Fig. S11(e). The obtained ARPES spectra clearly demonstrate that the ΓKM mirror plane can partition the band structure, indicating the mirror-symmetry protected nature of the electronic states in this material.